\documentclass[runningheads]{llncs}
\usepackage{graphicx}
\usepackage[colorlinks = true, allcolors=blue]{hyperref}
\usepackage{wrapfig}

\usepackage{booktabs}
\usepackage{multirow}
\usepackage{graphicx}
\usepackage[table,xcdraw]{xcolor}
\usepackage{subcaption}
\usepackage{caption}
\usepackage{subcaption}

\usepackage{microtype}
\usepackage[normalem]{ulem}
\useunder{\uline}{\ul}{}

\graphicspath{{./figs/}}
\begin{document}

\title{Improving Vaccine Stance Detection by Combining Online and Offline Data}
\titlerunning{CovExplain: Online/Offline Vaccine Stance Detection}
%
%
\author{Anique Tahir\inst{1} \and 
Lu Cheng\inst{1\and 2}  \and 
Paras Sheth\inst{1} \and 
Huan Liu\inst{1} }
\institute{Arizona State University, Tempe AZ 85281, USA \and University of Illinois Chicago, Chicago, Illinois, USA \\
    \email{\{artahir, lcheng35, psheth5, huanliu\}@asu.edu}}
%
\authorrunning{A. Tahir et al.}
%
%
\maketitle              

\begin{abstract}
Differing opinions about COVID-19 have led to various online discourses regarding vaccines. Due to the detrimental effects and the scale of the COVID-19 pandemic, detecting vaccine stance has become especially important and is attracting increasing attention. Communication during the pandemic is typically done via online and offline sources, which provide two complementary avenues for detecting vaccine stance. Therefore, this paper aims to (1) study the importance of integrating online and offline data to vaccine stance detection; and (2) identify the critical online and offline attributes that influence an individual's vaccine stance. We model vaccine hesitancy as a surrogate for identifying the importance of online and offline factors. With the aid of explainable AI and combinatorial analysis, we conclude that both online and offline factors help predict vaccine stance.
\end{abstract}
\section{Introduction}
In 2019, the COVID-19 pandemic started spreading across the globe, causing devastating consequences: millions of people dying from the disease, businesses shutting down, and people working from home. The pandemic introduced a major shift in people's lifestyles whose repercussions are observable to this day. Considering the rapid vaccine development, some people doubt the efficacy of the COVID-19 vaccine. Public opinions about vaccines in media can be roughly divided into anti-vaccine, pro-vaccine, and vaccine-hesitant. The \textbf{first} goal of this work is to effectively detect the vaccine stance, mainly whether a piece of information (e.g., tweets) shows support or opposition of the COVID-19 vaccine\footnote{Due to difficulty to accurately label if a tweet is vaccine-hesitant, we only consider pro- and anti-vaccine.}. 

The COVID-19 viral pandemic is an unprecedented global phenomenon and a personal experience with wide-range effects \cite{jackson2021global}. Accordingly, detecting a stance toward the COVID-19 vaccine may present unique challenges. For example, during this emergency, communication is typically done via numerous sources, such as newspapers, radio, television, and social media. The public opinions regarding the vaccine can be influenced by different voices, e.g., social media influencers, reporters from news media, and scientific and medical professionals. Therefore, social media data alone may be insufficient for accurate vaccine stance detection. One potential solution is to integrate both online (e.g., social media data) and offline data (e.g., demographics and policies) to help us understand the perspectives from both worlds. For example, knowing the State information of a user is helpful for detection due to its high correlation with the political parties.

However, linking online and offline data faces an inherent challenge because it requires us to understand the relation between offline information and the data shared by people on social media or news portals. Furthermore, it is essential to establish such a connection to understand the public opinion about the COVID-19 vaccine. Some public opinions, for instance, may be guided by misinformation or misconceptions. Our \textbf{second} goal is, therefore, to identify the essential online and offline attributes responsible for vaccine stance detection. To achieve these goals, we leverage advanced machine learning models for COVID-19 vaccine stance detection, and explainable AI approaches to analyze the predictions results on \textit{emerging} samples. Our contributions are summarized as follows:
\begin{itemize}
    \item We study the challenging problem of investigating if offline data can be used in connection with online data such as a user's social media discourse for improving vaccine stance detection.
    \item We propose a principled way to understand which online and offline attributes are helpful for vaccine stance detection at multiple levels using explainable AI approaches and combinatorial analysis.
    \item We experiment with real-world Twitter and offline data to provide practical insights for researchers and practitioners.
\end{itemize}


\section{Related Work}
\textbf{{Establishing a relationship between online and offline data}}
With the COVID-19 pandemic at large, information about the COVID-19 vaccines emerges from multiple sources, including online and offline sources. There have been increasing efforts to establish a link between online and offline data sources related to COVID-19. Establishing a connection between the online and offline data can generate better insights into understanding user behavior related to COVID-19. For instance, the authors in~\cite{feng2021integrating} aimed to study the relationship between policy developments and sentiment analysis for a few major COVID-19 related topics. Prior work~\cite{pfetsch2013critical} tried to establish a linkage between online communication and traditional mass media and discuss how it could aid in answering important questions such as how each media influences the other. Another line of work~\cite{moon2021determinants} aimed at studying the relationship between online and offline user shopping behavior during the COVID-19 pandemic. Some prior work also utilized real-world survey data to establish a link between online and offline user interactions in regards to the vaccine hesitancy~\cite{lee2022direct}. Although effective, surveys are able to analyze a very small population of users. To develop a better understanding of the online-offline relationship it is important to leverage a wider population which may not be feasible when utilizing surveys.
\\
\textbf{COVID-19 Vaccine Hesitancy Stance Detection}
While COVID-19 vaccines were fast developed, people have been showing hesitant attitudes toward the vaccine. Prior work focuses on identifying such individuals. For instance, the authors in~\cite{nyawa2022covid} aimed to evaluate the performance of different machine learning models and deep learning methods in identifying vaccine-hesitant tweets published during the COVID-19 pandemic. Another line of work~\cite{cotfas2021covid} aimed at studying the people's opinions in the UK after the first month of the release of the COVID-19 vaccines. Similarly, another work~\cite{masourislockdown} detected the stance of users in the Netherlands related to the lockdown policies. Given the dominance of language models such as BERT~\cite{devlin2018bert} in NLP tasks, various works have focused on investigating the use of BERT models for stance detection. For instance, the authors in~\cite{alibaevaanalyzing} aimed at leveraging BERT to detect COVID-19 related stance in Russian users. Prior work~\cite{glandt2021stance} provided a COVID-19 stance detection dataset and utilized self-training domain adaptation approaches to help improve the performance. The authors in~\cite{hou2022covmis} leveraged tweets to detect the stance of users towards COVID-19 related misinformation. 

We complement prior works by leveraging both online and offline information sources for the stance detection task, instead of relying on online content alone. We further investigate and identify the online and offline features critical for individual's vaccine stance detection to better understand the mechanism underlying COVID-19 vaccine hesitancy.

\section{Data Description}
\begin{figure}
     \centering
     \begin{subfigure}[b]{0.3\textwidth}
         \centering
         \includegraphics[width=\textwidth]{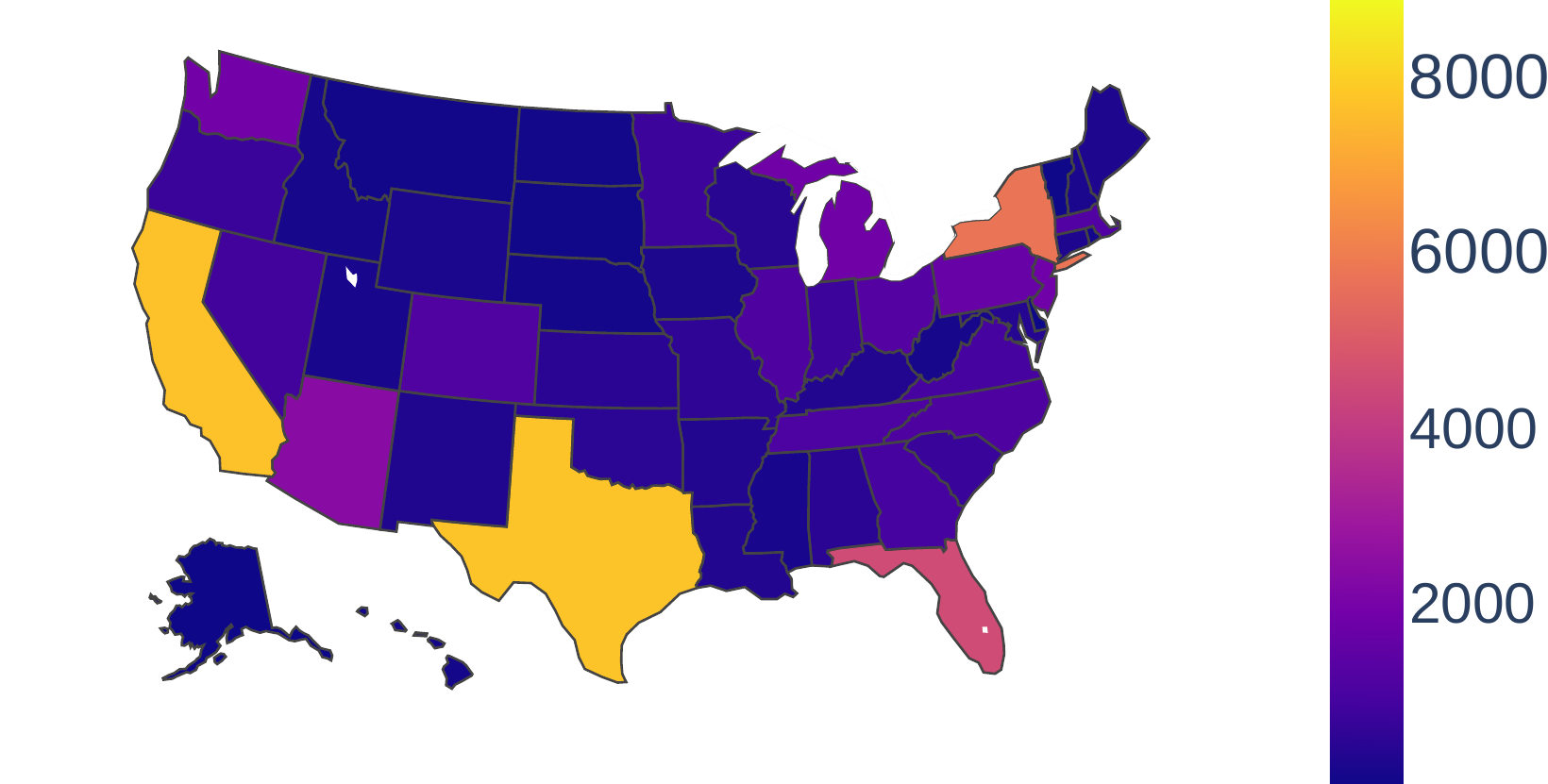}
         \caption{anti-vaccine}
     \end{subfigure}
     \hfill
     \begin{subfigure}[b]{0.3\textwidth}
         \centering
         \includegraphics[width=\textwidth]{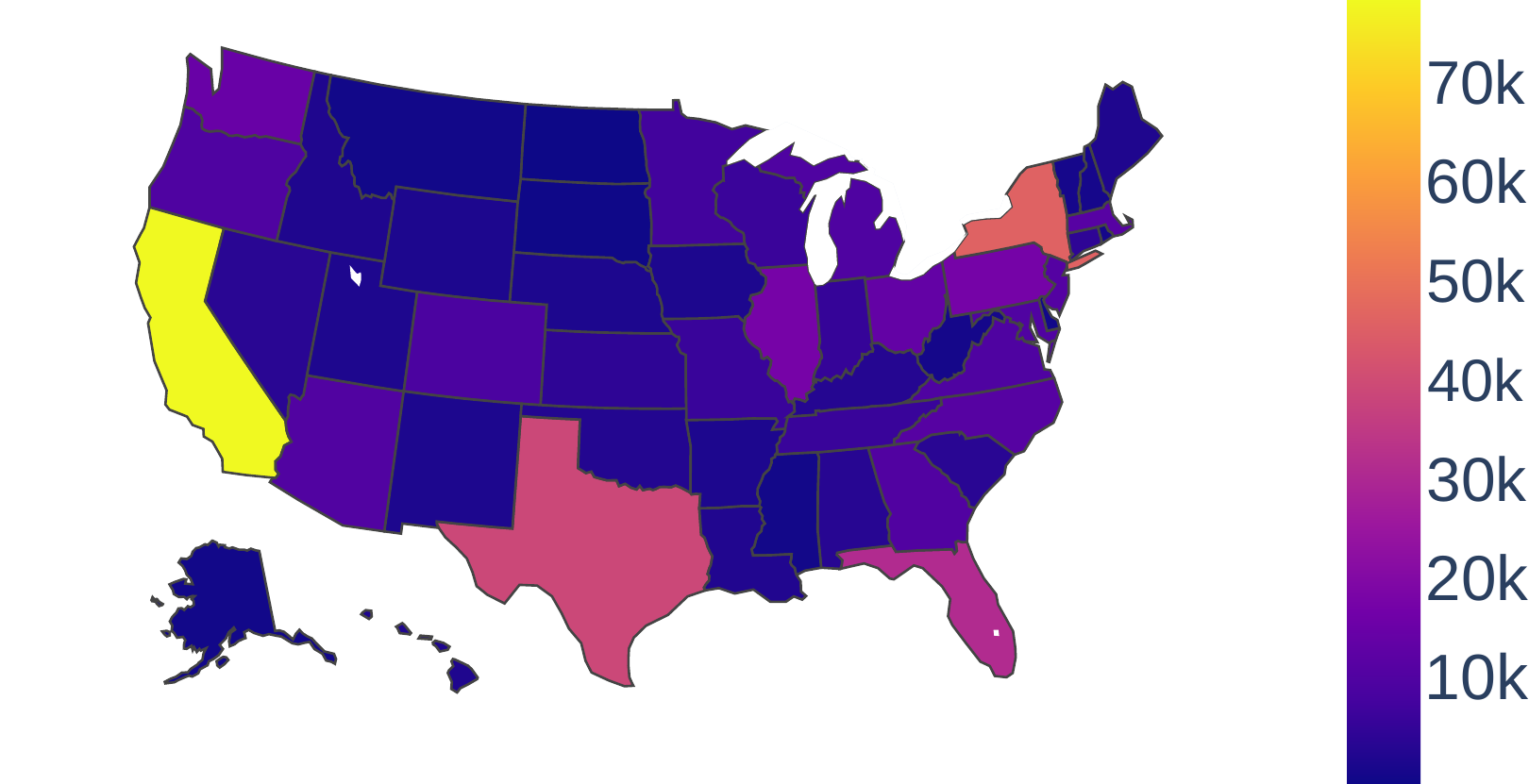}
         \caption{pro-vaccine}
     \end{subfigure}
     \hfill
     \begin{subfigure}[b]{0.3\textwidth}
         \centering
         \includegraphics[width=\textwidth]{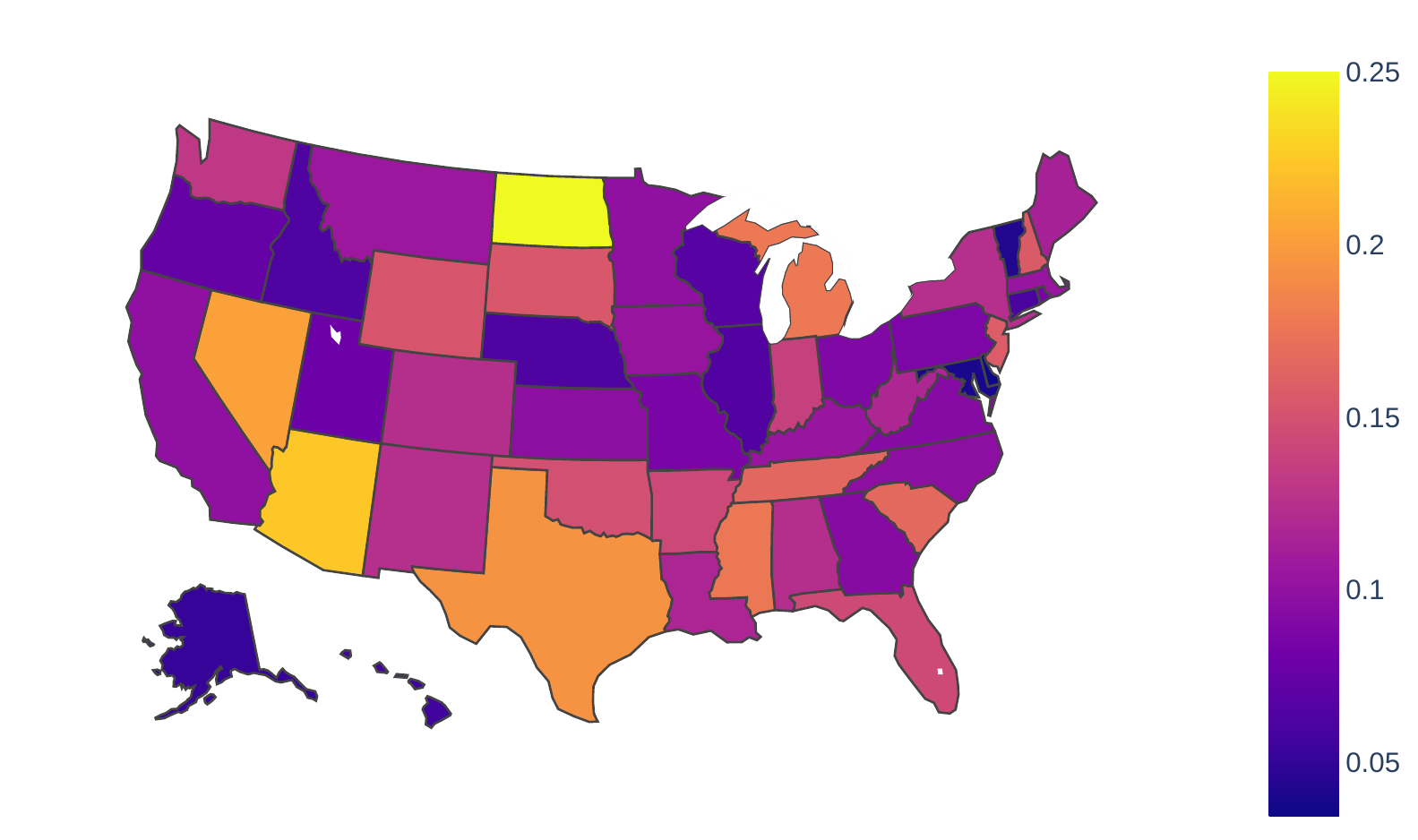}
         \caption{ratio}
     \end{subfigure}
     \caption{Choropleth maps showing the overall (a) anti-vaccine, (b) pro-vaccine, and (c) ratio distribution of the tweets we filtered for our analysis.}
     \label{fig:chloropleth}
\end{figure}
Data used for analysis is from CovaxNet\footnote{\url{https://github.com/jiangbohan/CoVaxNet}}, an Online-Offline data repository for COVID-19 vaccine hesitancy research~\cite{jiang2022covaxnet}. Mainly, Online data was collected from Twitter between 1/1/2021 and 1/1/2022. To obtain data on anti-vaccine and pro-vaccine users, 28 and 25 hashtags were used as filters, respectively. Examples include \textit{\#NoVaccine}, \textit{\#VaccinesKill}, \textit{\#GetVaccinated}, and \textit{\#VaccinesWork}. The collected tweets were then augmented with additional information such as race and gender. Geo-location from the user profile data was extracted and filtered out non-U.S. based users. As mentioned earlier, the user's geo-location is highly correlated with political beliefs. Thus, to analyze the role of the location on vaccine stance, we focus on the U.S. for this paper. The State information was also collected from Twitter user profiles. Finally, we obtained 630,009 tweets with 69,028 anti-vaccine and 560,981 pro-vaccine. Fig.~\ref{fig:chloropleth} shows the number of anti and pro-vaccine tweets across different states. Offline data were collected from U.S. Census Bureau Data, Government Response, and other sources. Please refer to the data repository for more details.
\begin{wrapfigure}{r}{0.5\textwidth}
    \vspace{-0.5cm}
    \begin{center}
    \includegraphics[width=0.47\textwidth]{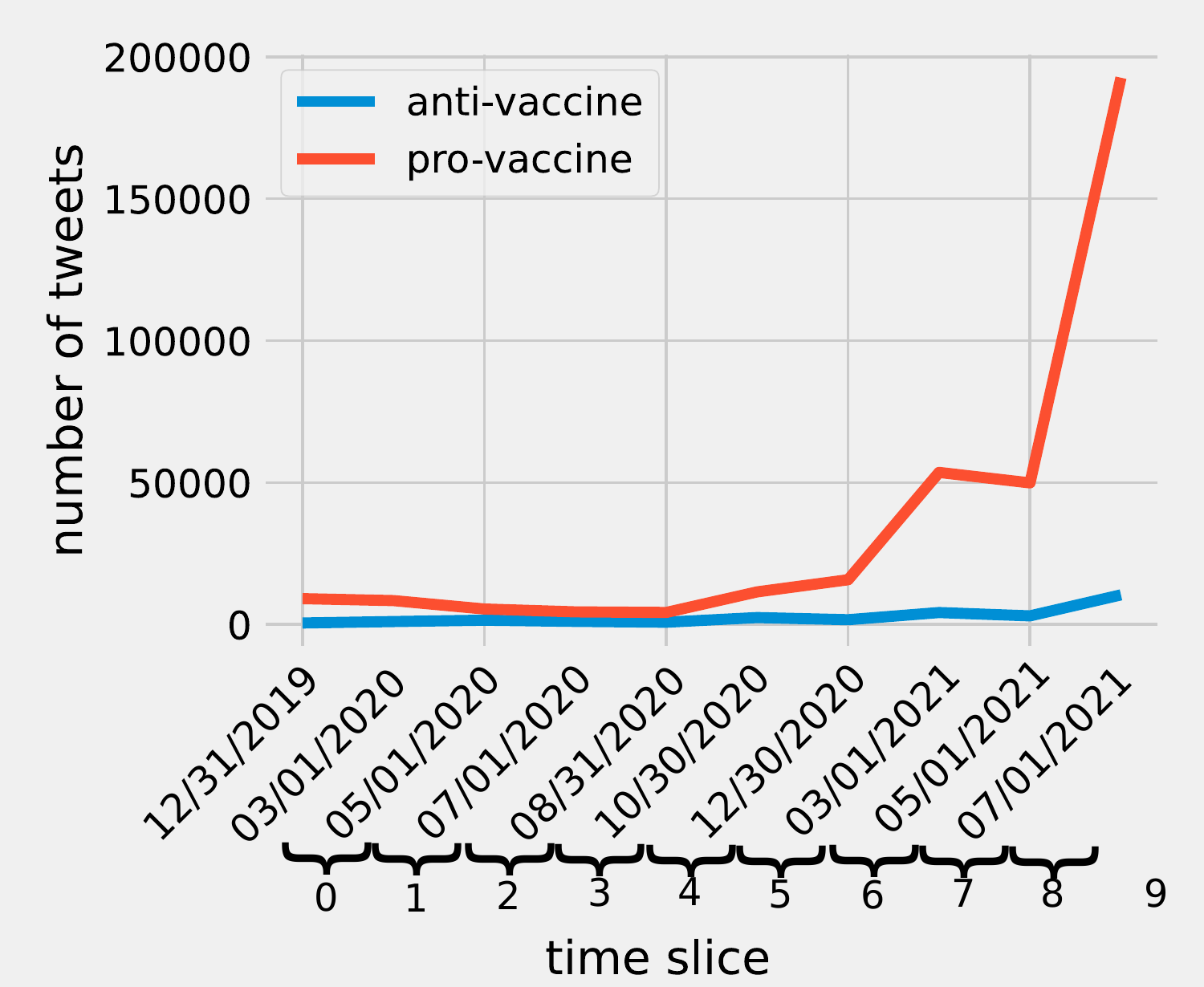}
    \end{center}
    \caption{The number of pro-vaccine and anti-vaccine tweets in each time slice.}
    \label{fig:data_tweet_numbers}
\end{wrapfigure}
\raggedbottom

Connecting the online and offline data is challenging due to the modality and granularity of different types of information. To tackle the problem, we extract users' demographic information (e.g., race and gender) from the Twitter user profile and offline data sources where possible. 
In \textit{chronological order}, we further divide the Twitter data into ten-time slices for the training and test splits. The test split is the last time slice. The reason for this splitting strategy is to evaluate whether our model has learned to generalize. Fig.~\ref{fig:data_tweet_numbers} shows the evolution of anti-vaccine and pro-vaccine tweets across time. We observe that (1) the pro-vaccine tweets heavily outnumber the anti-vaccine tweets over time (Fig.~\ref{fig:data_tweet_numbers}), and (2) California and Texas have the most significant number of anti-vaccine tweets while California and New York have the most significant number of pro-vaccine tweets (Fig.~\ref{fig:chloropleth}). (3) The gender distributions for anti and pro-vaccine groups are similar. Since we expect our model to learn to distinguish between the pro- and anti-vaccine individuals, we expect it to disregard the fact that some states are populous. 

\section{Method}
With the development of vaccines, there was a rapid generation of online data, e.g., tweets and posts, and offline data, e.g., news and COVID-19 statistics expressing people's opinions. As a result, people are divided into groups based on whether they support (pro-vaccine) or against (anti-vaccine) vaccines. However, this phenomenon also raised various questions, such as whether the online data in its entirety is sufficient to detect an individual's stance or whether the offline data influence the stance. Further, what features play a pivotal role in detecting an individual's vaccine stance. To tackle these questions, we propose a principled framework~(\textit{CovExplain}) that leverages online and offline attributes to detect an individual's stance. Importantly, \textit{CovExplain} explains which online and offline features are helpful for vaccine stance detection.

The modus operandi of analyzing such social media text is to extract the semantics using a large language model. However, aside from online information, including the tweet content and profile description, a user's auxiliary offline data such as their race, gender, and state may hold additional cues influencing the user's stance. To this end, we hypothesize that adding and removing users' demographic and other auxiliary information can help identify the stance of the users. CovExplain is our pipeline for extracting user-based information from raw social media data. It combines all the user and semantic online and offline information from our pipeline to train a model capable of identifying the users' stance. Finally, we employ intervention analysis and explanation frameworks to explore the relative importance of different user-based features and the context in which the discourse takes place.

In the following parts of this section, we illustrate our approach in two steps. First, we introduce the preprocessing module and then discuss the classification model to identify an individual's stance.



\subsection{Data Preprocessing and Model Generation}

Tweets often consist of hashtags and URLs along with the text. Since vaccine stance was labeled based on the hashtag search, we replaced all the hashtags and URLs in tweets during the model training and testing. We extract features using Covid-Twitter-bert~\cite{muller2020covid}. The descriptions in the users' profiles are relatively different in terms of syntax compared to the tweets. Thus, we use RoBERTa~\cite{liu2019roberta} model to get the description features.

Apart from the users' online tweet content, we leverage the users' offline profile attributes for vaccine stance detection. To understand how the demographic features affect a user's vaccine stance, we utilize the user's state, race (Race), race inferred from profile picture ($Race_{pic}$), and gender for the offline dataset. The offline features are categorical variables and encoded using one-hot representation. Formally, given a categorical feature $c \in C$, we define its one-hot embedding as:
\begin{equation}
    \mathcal{O}(c) = \sum_{i=1}^m \mathbf{1}\{c = C_i\} \mathbf{v}_i, \label{eq:one-hot}
\end{equation}
where $m$ is the number of categories, $C=\{c_1, ..., c_m\}$, and $\mathbf{v}_i$ is the $i$th standard basis vector. We define the one-hot representation of a categorical feature $c$ with $m$ categories as the $m$-dimensional vector $[\mathcal{O}(c)]$. Following a traditional setting, we sample an equal number of pro-and anti-vaccine tweets to construct a balanced dataset to minimize the effect of bias in the analysis.

Online and offline datasets consist of a variety of multi-modal features. To analyze the contributing features toward vaccine stance from both datasets, we leverage functions that facilitate the addition or removal of features and manufacture a model that fits our dynamic features' dimensions. Formally, let $n$ be the number of features and $\phi$ represent the model parameters. We define a function $\tau(\cdot)$ that initializes $\phi$, i.e., $\phi_n = \tau(n)$. The prediction function $f(\cdot)$ infers the logits $\hat{y} = f(x, \phi_n)$, where $x$ represent the input features.

We use BERT~\cite{devlin2018bert} to extract the text-based features of the tweets and RoBERTa~\cite{liu2019roberta} for extracting the features from user profile descriptions as they are relatively different in terms of syntax and format. We freeze the model parameters and remove the classification heads and the pooling layers. The last four hidden layers of the language models are used as the features for further classification. Let $h_i$ be the last $i$'th hidden layer of BERT, and $f(\cdot)$ be the classification function with parameters $\phi$. Let $\mathcal{L}$ be a loss function and $y$ be the ground truth labels. The objective for fine-tuning the language models can be defined as:
\begin{equation}
    \arg\!\min\limits_{\phi} \mathcal{L}(y, f(h_1 \circ h_2 \circ h_3 \circ h_4, \phi)),
\end{equation}
where $\circ$ denotes concatenation.

\subsection{Stance Detection Model}
After the data preprocessing, we have (1) the online features, including the tweet text and users' profile descriptions, and (2) the offline features, including the state, race, and gender information.

A simple multi-layer perceptron~(MLP) with six layers is used with the semantic features of the tweets and description combined with the one-hot representations of the offline features. The layers of the MLP have batch normalization, LeakyRELU~\cite{zhang2017dilated} activation, and Monte Carlo dropout~\cite{srivastava2014dropout}. We use softmax cross entropy loss for optimizing the network:
\begin{equation}
    \hat{y}_{i} = \sigma(z)_i = \frac{e^{z_i}}{\sum_{j=1}^K e^{z_j}},
\end{equation}
where $z$ is the softmax feature vector, $z_i$ is the $i$th element of $z$, and $K = |z|$ is the cardinality of our feature vector. A binary cross-entropy loss function is used to train the model:
\begin{equation}
    \mathcal{L}(y, \hat{y}) = - \sum_{i=1}^N y_i \log(\hat{y}_i) + (1 - y_i) \log(1 - \hat{y}_i),
\end{equation}
where $y$ is the one-hot label vector and $y_i$ is the $i$th element of the $y$ vector, $y_i \in \{0, 1\}$, and $N = |y| = |\sigma(z)|$ and $\hat{y} = \{\sigma(z)_1, ..., \sigma(z)_N\}$.

\subsection{Explanation Methodology}
We investigate Twitter users' vaccine stance using local and global explanations. Local explanations are based on our observations for specific instances in the dataset, e.g., tweet, description, or the weight of a particular feature towards prediction. For local explanations, we use Shapley~\cite{winter2002shapley} values. SHAP (SHapley Additive exPlanations)~\cite{NIPS2017_7062} is an explanatory framework based on game theory principles. Optimal credit allocation is used for generating local explanations. SHAP can be used to explain instances provided to an NLP model by highlighting the importance of each word's contribution to the prediction. In essence, the explanation framework allows us to pinpoint the componets of a sample which, if removed, would change the model's predictions.

We use combinatorial analysis over various machine learning models to get explanations at a more global scale. Specifically, we divide our data into Online and Offline features. Next, we use all sets of offline features with a cardinality of three to train the models. Finally, we look at models which train on a set of all offline and online features separately.



\section{Experimental Results}

We aim to answer the following research questions: \textbf{RQ. 1} Does combining online, and offline data help improve the accuracy of vaccine stance detection? And \textbf{RQ. 2} What features contribute to the most improvement of the performance? 

\subsection{Experimental Setup}
Our classification model uses PyTorch~\cite{NEURIPS2019_9015} for implementation. We use the covid-twitter-bert and roberta-large pre-trained models and fine-tune them using the customed six-layer Batch Normalized MLP with hidden dimensions of 1024 neurons and a dropout layer with a drop probability of 0.2. We train the model using the AdamW~\cite{loshchilov2018fixing} optimizer with a learning rate of 1e-2. The model is trained for 80 epochs with a batch size of 256 samples. 

We use standard machine learning models as our baselines. Linear regression is used to identify the linear relations between the features for comparison. Gaussian Naive Bayes is used to find probabilistic relations between the features and the stance. Finally, Support Vector Machine (SVM) with Radial Bias Function (RBF) kernel is used for finding a non-linear decision boundary for the features. To ensure there is no information leakage from the future time steps, we split the dataset chronologically into 10 time slices where the last slice is used for evaluation. 


\begin{table}[]
\centering
\resizebox{\columnwidth}{!}{%
\begin{tabular}{@{}llcccc@{}}
\toprule
                          &                                             & \textbf{CovExplain(Ours)}                         & \textbf{Linear}                      & \textbf{Naive Bayes}                 & \textbf{SVM}                         \\ \midrule
                          & \cellcolor[HTML]{EFEFEF}Tweets                                      & \cellcolor[HTML]{EFEFEF}73.16 $\pm$ 4.8                         & \cellcolor[HTML]{EFEFEF}70.41 $\pm$ 4.1                         & \cellcolor[HTML]{EFEFEF}52.95 $\pm$ 4.75                        & \cellcolor[HTML]{EFEFEF}{\ul 52.62 $\pm$ 2.0}                   \\
\multirow{-3}{*}{Online}  & Description         & 78.95 $\pm$ 4.1 & 68.62 $\pm$ 5.3 & 52.29 $\pm$ 4.6 & 52.33 $\pm$ 2.8 \\ 
  & \cellcolor[HTML]{EFEFEF}All              & \cellcolor[HTML]{EFEFEF}{\ul 81.00 $\pm$ 3.8}                   & \cellcolor[HTML]{EFEFEF}{\ul 74.70 $\pm$ 3.5}                   & \cellcolor[HTML]{EFEFEF}55.58 $\pm$ 4.58                        & \cellcolor[HTML]{EFEFEF}51.58 $\pm$ 3.2                         \\

\cmidrule(r){1-1}
                          & State+Race+$Race_{pic}$                        & 67.08 $\pm$ 3.8                         & 68.08 $\pm$ 3.3                         & {\ul 68.16 $\pm$ 3.0}                   & 50.70 $\pm$ 4.5                         \\
                          & \cellcolor[HTML]{EFEFEF}State+Race+Gender   & \cellcolor[HTML]{EFEFEF}67.91 $\pm$ 3.4 & \cellcolor[HTML]{EFEFEF}69.00 $\pm$ 2.8 & \cellcolor[HTML]{EFEFEF}67.91 $\pm$ 3.2 & \cellcolor[HTML]{EFEFEF}47.83 $\pm$ 2.8 \\
                          & State+$Race_{pic}$+Gender                        & 59.12 $\pm$ 4.9                         & 56.12 $\pm$ 4.2                         & 58.04 $\pm$ 4.1                         & 48.37 $\pm$ 4.4                         \\
\multirow{-5}{*}{Offline} & \cellcolor[HTML]{EFEFEF}Race+$Race_{pic}$+Gender & \cellcolor[HTML]{EFEFEF}66.45 $\pm$ 3.2 & \cellcolor[HTML]{EFEFEF}67.33 $\pm$ 2.9 & \cellcolor[HTML]{EFEFEF}67.33 $\pm$ 2.6 & \cellcolor[HTML]{EFEFEF}50.04 $\pm$ 0.6 \\ 
                          & All             & 67.04 $\pm$ 3.8 & 66.29 $\pm$ 3.1 & 67.41 $\pm$ 2.8 & 50.37 $\pm$ 0.8 \\

\cmidrule(r){1-1}

\multirow{-1}{*}{Hybrid}  & \cellcolor[HTML]{EFEFEF}Online+Offline                              & \cellcolor[HTML]{EFEFEF}\textbf{83.37 $\pm$ 3.3}                & \cellcolor[HTML]{EFEFEF}\textbf{79.04 $\pm$ 3.7}                & \cellcolor[HTML]{EFEFEF}\textbf{68.72 $\pm$ 3.0}                & \cellcolor[HTML]{EFEFEF}\textbf{53.27 $\pm$ 1.8}                \\ \bottomrule
\end{tabular}%
}
\caption{Classification results of different fine-tuning models showing mean and standard deviation over 20 random samples. Bold and underlined results represent the highest performance scores and the runner-up for each model, respectively. Offline features are represented by the user demographics and the online features are encapsulated in tweets and user profile description.}
\label{tbl_results}
\end{table}

\subsection{Classification Results}

We perform combinatorial analysis by training various models on different subsets of features. We first train all the compared models on online and offline features separately. We utilize the \textit{Tweets}, and users' profile biography i.e. \textit{Description} as online features. We use the user's \textit{State}, \textit{Race}, \textit{Gender}, and \textit{$Race_{pic}$} as offline features. $Race_{pic}$ is the DeepFace race detection, while Race is extracted from the name of the user. We use both features since there is a chance of misclassification from both methods of race extraction. When exclusively using offline features, we consider various combinations of offline features (e.g., \textit{State + Race + Gender} and \textit{State + Race + $Race_{pic}$}). We then integrate online and offline features and refer to it as \textit{Hybrid}. Results are presented in Table.~\ref{tbl_results}.

We make the following observations: (1) Online features have overall better predictive capability compared to the offline features themselves. We believe this is because the offline features have less expressiveness than online features such as tweets, which can represent vast information. (2) We also observe that the users' profile biography (i.e. `Description') can be equally as informative as their tweets. This is a remarkable observation since the tweets in the dataset were collected based on the condition that they are related to COVID-19. However, the users' profile description does not necessarily meet this criterion. (3) Our hybrid approach can achieve improved performance by combining the online and offline features.
\begin{figure}[h]
    \begin{center}
         \begin{subfigure}[b]{0.8\textwidth}
         \centering
         \includegraphics[width=\textwidth]{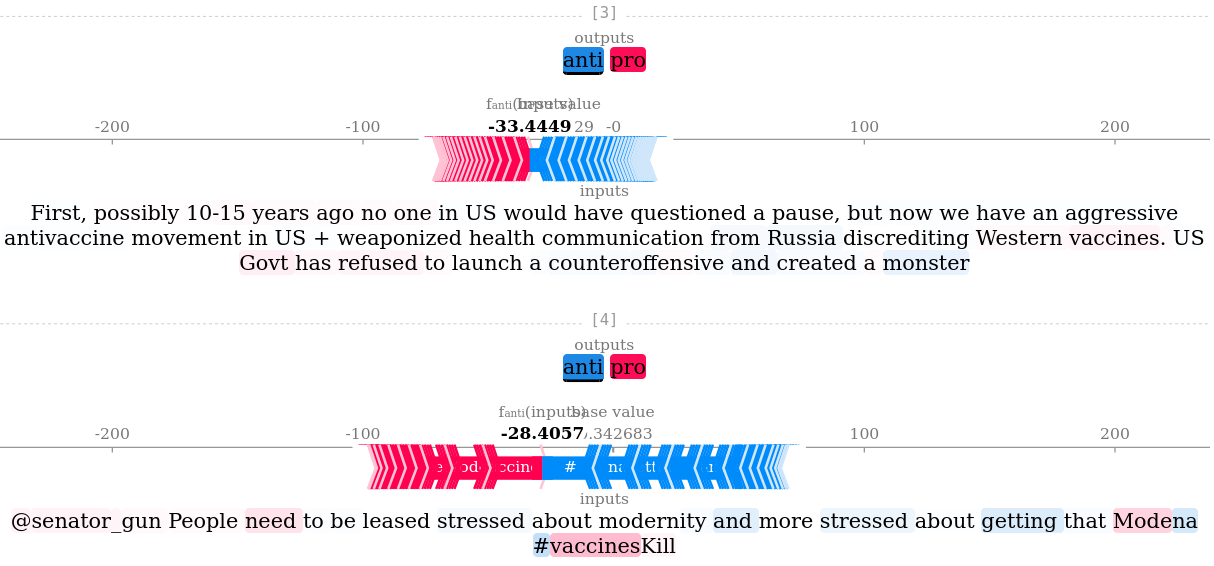}
         \caption{anti-vaccine Shapely values per word for sample tweets}
     \end{subfigure}
          \begin{subfigure}[b]{0.8\textwidth}
         \centering
         \includegraphics[width=\textwidth]{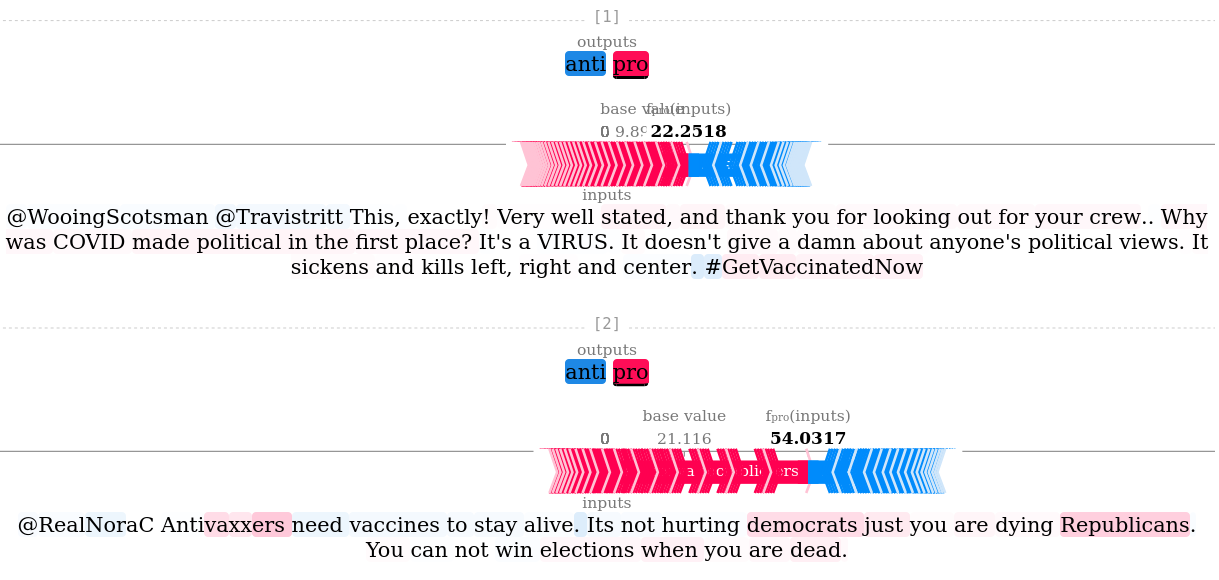}
         \caption{pro-vaccine Shapely values per word for sample tweets}
     \end{subfigure}
    \caption{Shapely additive explanations for the Text based classifier. The Shapely values are computed on the tweet content, indicating which words are predictive of the user's vaccine stance.}
    \label{fig:shap_anti}
    \end{center}
\end{figure}
\raggedbottom
Our combinatorial analysis shows that race and gender are the most contributing factors among the offline features of predicting vaccine stance. In addition, these offline features outperform other offline feature combinations used for our analysis. However, we observe that using all offline features adds spurious correlations from the training data, leading to decreased testing performance. These results answer \textbf{RQ. 1}, showing that combining the online and offline data significantly improves the prediction accuracy of the different models.

\subsection{Explanation Results}
We use SHAP values to answer \textbf{RQ. 2}. Here, for SHAP analysis we focus on online features as they are more expressive than offline features. Fig.~\ref{fig:shap_anti} shows a sample of our SHAP explanations to pro- and anti-vaccine tweets. We find that the discourse by individuals with anti-vaccine beliefs is aimed toward entities such as companies and countries, e.g., the first tweet Fig.~\ref{fig:shap_anti}(a) shows the importance of the word `Russia'. In contrast, the second tweet in Fig~\ref{fig:shap_anti}(b) shows the importance of `vaxxers'. On the other hand, pro-vaccine individuals' discourse is targeted toward anti-vaccine individuals and toward spreading vaccine awareness. Anti-Vaccine tweets are more politically driven, while Pro-Vaccine tweets are more emotionally driven. 
Since the distribution of gender is similar for both the anti- and pro-vaccine groups, it does not provide enough discriminating information for classification by itself. However, the performance increases when gender is combined with attributes such as race and state. 

\section{Conclusion}
Due to the potential of connecting online and offline data to help vaccine stance detection, we develop a framework to model and explain vaccine stances. We use our model as a surrogate for explaining the factors responsible for individuals' stances. 
We show that online and offline factors can be combined to achieve better predictive performance for vaccine stance. We analyze Twitter users' posts to identify beliefs responsible for their stance.
There are several interesting future directions. Additional offline data such as user-level policies might help in improving accuracy since our analysis suggests that anti-vaccine individuals are interested in the government. Additionally, our model can be used to enhance vaccine awareness.

\subsubsection*{Acknowledgments}
The research for this paper was supported by the Office of Naval Research (ONR) under grant N00014-21-1-4002.

%
%
%
\bibliographystyle{splncs04}
\bibliography{main}
%
%
%
%
%
\end{document}